\documentclass[aps, preprint, superscriptaddress]{revtex4-2}
\usepackage{graphicx}
\usepackage{makecell, multirow}
\usepackage{color, bm, soul, xcolor}
\usepackage{amsmath, amssymb}
\usepackage{gensymb}
\usepackage{geometry}
\usepackage{braket}
\usepackage[colorlinks=true, linkcolor=blue, citecolor=blue, urlcolor=blue]{hyperref}
\newcommand{\eg}{{\em e.\,g.}}
\newcommand{\Ec}{{$\mathcal{E}_{\rm c}$}}
\newcommand{\Es}{{$\mathcal{E}_{\rm s}$}}
\newcommand{\Hfi}{{Hf$_{\rm i}$}}
\newcommand{\Sii}{{Si$_{\rm i}$}}

\begin{document}

\title{Origin of Interstitial Doping Induced Coercive Field Reduction in Ferroelectric Hafnia}
\author{Tianyuan Zhu}
\thanks{These two authors contributed equally.}
\affiliation{Department of Physics, School of Science, Westlake University, Hangzhou, Zhejiang 310030, China}
\affiliation{Institute of Natural Sciences, Westlake Institute for Advanced Study, Hangzhou, Zhejiang 310024, China}

\author{Liyang Ma}
\thanks{These two authors contributed equally.}
\affiliation{Department of Physics, School of Science, Westlake University, Hangzhou, Zhejiang 310030, China}

\author{Xu Duan}
\affiliation{School of Science, Zhejiang University of Science and Technology, Hangzhou, Zhejiang 310023, China}

\author{Shi Liu}
\email{liushi@westlake.edu.cn}
\affiliation{Department of Physics, School of Science, Westlake University, Hangzhou, Zhejiang 310030, China}
\affiliation{Institute of Natural Sciences, Westlake Institute for Advanced Study, Hangzhou, Zhejiang 310024, China}

\begin{abstract}
{Hafnia-based ferroelectrics hold promise for nonvolatile ferroelectric memory devices. However, the high coercive field required for polarization switching remains a prime obstacle to their practical applications. A notable reduction in coercive field has been achieved in ferroelectric Hf(Zr)$_{1+x}$O$_2$ films with interstitial Hf(Zr) dopants~[\href{https://www.science.org/doi/abs/10.1126/science.adf6137}{Science \textbf{381}, 558 (2023)}], suggesting a less-explored strategy for coercive field optimization. Supported by density functional theory calculations, we demonstrate the $Pca2_1$ phase, with a moderate concentration of interstitial Hf dopants, serves as a minimal model to explain the experimental observations, rather than the originally assumed rhombohedral phase. Large-scale deep potential molecular dynamics simulations suggest that interstitial defects promote the polarization reversal by facilitating $Pbcn$-like mobile 180$\degree$ domain walls. A simple pre-poling treatment could reduce the switching field to less than 1 MV/cm and enable switching on a subnanosecond timescale. High-throughput calculations reveal a negative correlation between the switching barrier and dopant size and identify a few promising interstitial dopants for coercive field reduction.}
\end{abstract}

\maketitle

\clearpage
Ferroelectric hafnia (HfO$_2$) has emerged as a promising candidate for integrating ferroelectric functionalities into integrated circuits, enabled by its robust nanoscale ferroelectricity and exceptional silicon compatibility~\cite{Boscke11p102903,Cheema20p478,Schroeder22p653}. However, the high coercive field (\Ec) required to switch the polarization in HfO$_2$-based thin films remains a critical issue, impeding the commercialization of this silicon-compatible ferroelectric~\cite{Schroeder22p653,Noheda23p562}. Typically, polycrystalline thin films of hafnia fabricated through atomic-layer deposition display an {\Ec} exceeding 1 MV/cm~\cite{Schroeder14p08LE02}. High-quality epitaxial thin films, obtained by pulsed laser deposition, can show even higher {\Ec} values of $\approx$ 2--5 MV/cm~\cite{Wei18p1095,Lyu19p220,Estandia19p1449,Song20p3221}. The need for applying high electric fields, close to the material's breakdown strength, for polarization reversal seriously limits its field cycling endurance. 
Reducing \Ec~without sacrificing the nanoscale ferroelectricity of hafnia is a pressing challenge for the wider adoption of HfO$_2$-based ferroelectrics.

Various strategies such as doping~\cite{Qi20p214108,Yang20p064012}, strain~\cite{Wei22p154101,Zhou22peadd5953}, and superlattices~\cite{Zhao24p256801} have been theoretically explored to lower the barrier for polarization switching, aiming to reduce \Ec. Investigations based on density functional theory (DFT) revealed that the polarization switching process in the ferroelectric othorhombic ($O$) $Pca2_1$ phase often involves an intermediate tetragonal ($T$) $P4_2/nmc$ phase, and the energy difference between these two phases could serve as a measure of the switching barrier~\cite{Yang20p064012, Ma23p256801}. Substitutional Si has been suggested as an effective dopant for reducing \Ec~due to its intrinsic $sp^3$ bonding with oxygen, which helps stabilize the intermediate $T$ phase~\cite{Yang20p064012, Falkowski18p73}. Conversely, most other substitutional dopants have shown limited effectiveness in influencing \Ec~based on DFT calculations~\cite{Yang20p064012}. Recently, a substantial reduction in \Ec, down to $\approx$0.65 MV/cm, has been achieved in thin films of Hf(Zr)$_{1+x}$O$_2$, a composition rich in hafnium-zirconium [Hf(Zr)]~\cite{Wang23p558}. This reduction was attributed to the interstitial doping of Hf and Zr atoms into a polar rhombohedral ($R$) $R3m$ phase. DFT calculations show that a 7\%-strained $R$-phase Hf$_{1.08}$O$_2$ exhibits a switching barrier of 7.6 meV/atom, significantly lower than that of 20 meV/atom in stoichiometric $R$-HfO$_2$.

However, assuming a polar $R$ phase in Hf$_{1.08}$O$_2$ films grown via magnetron sputtering leads to notable discrepancies between theory and experiment~\cite{Fina21p1530, Wei23preserchsquare}. The unstrained $R$ phase is actually a nonpolar cubic $P\bar{4}3m$ phase~\cite{Schroeder22p653, Zhu23pL060102}; the polar $R3m$ symmetry only emerges under an equibiaxial compressive strain within the (111) crystallographic plane. Although the relative stability between the $R$ and $O$ phases can be reversed with 8\% interstitial doping~\cite{Wang23p558}, highly-doped $R$-Hf$_{1.08}$O$_2$ without strain remains nonpolar; a giant compressive strain of 7\% is nevertheless required to induce a polarization of $\approx$15 $\mu$C/cm$^2$, still below the experimental value of 22 $\mu$C/cm$^2$ in polycrystalline films~\cite{Wang23p558}. Moreover, X-ray diffraction (XRD) patterns of Hf(Zr)$_{1+x}$O$_2$ thin films show only a slight offset from both unstrained $R$ and $O$ phases~\cite{Qi20p257603}, indicating the absence of high strains.

In this work, we combine DFT and deep potential molecular dynamics (DPMD) simulations~\cite{Zhang18p143001} to investigate how interstitial Hf doping reduces the coercive field in ferroelectric HfO$_2$. We find that interstitial Hf dopants diminish the energy difference between the polar orthorhombic $Pca2_1$ phase and the intermediate tetragonal $P4_2/nmc$ phase, lowering the switching barrier and coercive field. Unlike the $R$ phase, which demands additional compressive strains for polarization, the $Pca2_1$ phase of Hf$_{1.03}$O$_2$, with a moderate doping concentration of 3\%, already exhibits a low switching barrier and a high polarization. We propose that the $Pca2_1$ phase of HfO$_2$ with interstitial Hf defects better explains experimental observations. Large-scale MD simulations at finite temperatures with a DFT-derived machine learning force field demonstrate that 0.5\% interstitial doping in $Pca2_1$ HfO$_2$ is enough to reduce the switching field significantly. Moreover, after a pre-poling treatment during which interstitial dopants induce the formation of mobile domain walls, the switching field can drop below 1 MV/cm, comparable to experimental results. Further DFT calculations covering various interstitial dopants reveal a negative correlation between the switching barrier and dopant size. 

All DFT calculations are carried out using the Vienna \textit{ab initio} simulation package (VASP)~\cite{Kresse96p11169} with the projector augmented-wave (PAW) method~\cite{Blochl94p17953, Kresse99p1758} and the Perdew-Burke-Ernzerhof (PBE) exchange correlation functional~\cite{Perdew96p3865}. The cut-off energy for the plane-wave basis is set to 600 eV. The Brillouin zones of the 12-atom pseudocubic unit cell and the 2$\times$2$\times$2 supercell for doped systems are sampled by $\Gamma$-centered 4$\times$4$\times$4 and 2$\times$2$\times$2 Monkhorst-Pack~\cite{Monkhorst76p5188} $k$-point meshes, respectively. All the structures are fully optimized until the atomic forces converge to 0.01 eV/\AA. The polarization values are determined using the Berry phase method~\cite{KingSmith93p1651, Vanderbilt93p4442}. The polarization switching pathway is determined using the nudged elastic band (NEB) method~\cite{Sheppard12p074103}. The bond order analysis is performed with the 
crystal orbital Hamilton population (COHP) method~\cite{Dronskowski93p8617} implemented in LOBSTER~\cite{Maintz16p1030}. To investigate the ferroelectric switching in Hf$_{1+x}$O$_2$ at finite temperatures, we perform isobaric-isothermal ensemble ($NPT$) MD simulations using a deep neural network-based force field~\cite{Wu21p024108,Wu23p144102}, with model accuracy verification and sample inputs available via an online {\href{https://nb.bohrium.dp.tech/detail/1034100470}{notebook}}~\cite{notebookforL48}. The electric fields are included in MD simulations using the ``force method"~\cite{Umari02p157602, Liu16p360, Ma23p256801}, where an additional force $\mathcal{F}_i$ is added on the ion $i$ according to $\mathcal{F}_i = Z_i^*\cdot \mathcal{E}$, with $Z_i^*$ being the Born effective charge (BEC) tensor of ion $i$.

\begin{figure}[t]
\includegraphics[width=0.9\textwidth]{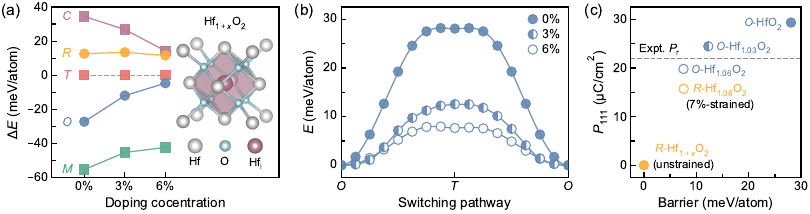}
\caption{Phase stability and ferroelectric switching of Hf$_{1+x}$O$_2$. (a) Energy landscape of monoclinic ($M$) $P2_1/c$, orthorhombic ($O$) $Pca2_1$, tetragonal ($T$) $P4_2/nmc$, rhombohedral ($R$) $R3m$, and cubic ($C$) $Fm\bar{3}m$ phases under interstitial doping of different concentrations (3\% and 6\% in short of 3.125\% and 6.25\%, respectively). The energy of $T$ phase is set to zero as a reference. The inset schematically shows a interstitial Hf atom intercalated within a void (purple polyhedron) formed by surrounding 6 Hf and 8 O atoms. (b) Energy variation along the polarization switching pathway of the $O$ phase under different doping concentrations. (c) Polarization projected along the [111] crystallographic orientation and switching barrier of the $O$ and $R$ phases. The results of 7\%-strained $R$ phase (hollow yellow circle) and experimental remanent polarization (dashed line) are extracted from Ref.~\cite{Wang23p558}.}
\label{fig_phase}
\end{figure}

We start with an analysis of the phase stability of Hf$_{1+x}$O$_2$ with interstitial Hf dopants (\Hfi), considering the monoclinic ($M$) $P2_1/c$, polar orthorhombic ($O$) $Pca2_1$, tetragonal ($T$) $P4_2/nmc$, rhombohedral ($R$) $R3m$, and cubic ($C$) $Fm\bar{3}m$ phases. Two doping concentrations, 3.125\% and 6.25\% (abbreviated as 3\% and 6\%), are modeled by intercalating one and two excess Hf atom within the hollow sites in a 96-atom 2$\times$2$\times$2 supercell, respectively. As shown in Fig.~\ref{fig_phase}(a), pristine HfO$_2$ polymorphs exhibit an inverse correlation between the thermodynamic stability and crystal symmetry.
With increasing \Hfi~concentration, the energies of these phases become closer, suggesting higher \Hfi~formation energy in lower-symmetry $M$ and $O$ phases but lower in higher-symmetry $C$ and $R$ phases. This dependence is due to different effective void sizes in each phase [see the inset of Fig.~\ref{fig_phase}(a)]. Specifically, the $M$ and $O$ phases have smaller voids due to alternating fourfold-coordinated (4C) and threefold-coordinated (3C) oxygen atoms, hindering \Hfi~integration and resulting in higher formation energy.

Consistent with previous DFT results~\cite{Wang23p558}, our calculations show that, compared to the $T$ phase, \Hfi~doping destabilizes the lower-symmetry $M$ and $O$ phases while imposing minimal impact on the relative stability of the $R$ phase. This suggests that the $R$ phase would become energetically more favorable than the $O$ phase above a critical \Hfi~concentration, if it remains stable. However, the \Hfi-doped $R$ phase is nonpolar, requiring a (111) in-plane compressive strain to achieve a polarized $R$ phase. Using a 36-atom hexagonal supercell with a 7\% in-plane compressive strain and an excess Hf atom, \citeauthor{Wang23p558} reported a polarized $R$-Hf$_{1.08}$O$_2$ with a low switching barrier of 7.6 meV/atom~\cite{Wang23p558}. Notably, such a high strain state should correspond to a large out-of-plane interplanar spacing, $d_{111}=3.11$ \AA. This does not align with the experimental XRD data of Hf$_{1+x}$O$_2$ films, which shows a (111) peak at $2\theta=30.1\degree$, indicating only a 0.1$\degree$ shift compared to the stoichiometric $O$ phase (with $d_{111}=$2.96 \AA, see Fig.~S1).

In contrast to the $R$ phase, the $O$ phase is inherently polar and does not depend on applied strain for polarization. The ferroelectric switching barrier in the $O$ phase can be estimated from the energy difference between the $T$ phase and $O$ phase ($\Delta E = E^T - E^O$)~\cite{Yang20p064012}. As illustrated in Fig.~\ref{fig_phase}(a), interstitial doping elevates the energy of the $O$ phase relative to the $T$ phase, thereby reducing $\Delta E$. 
Figure~\ref{fig_phase}(b) plots the NEB energy variation along the switching pathway in the $O$ phase for different doping concentrations. The switching barrier decreases drastically from 28 meV/atom in pristine HfO$_2$ to 12.3 meV/atom in Hf$_{1.03}$O$_2$, and further drops to 7.5 meV/atom in Hf$_{1.06}$O$_2$.

As summarized in Fig.~\ref{fig_phase}(c), for the $O$ phase, both the switching barrier and polarization decrease with increasing \Hfi~concentration. This reduction in polarization is attributed to the diminished local displacements of polar oxygen atoms surrounding \Hfi. Further increasing \Hfi~concentration destabilizes the $O$ phase, causing it to transform into a $Pbcn$-like structure during DFT geometry relaxation (see Fig.~S2). Importantly, $O$-Hf$_{1.06}$O$_2$, with a lower \Hfi~concentration, exhibits a switching barrier (7.5 meV/atom) comparable to that of 7\%-strained $R$-Hf$_{1.08}$O$_2$. It also features a higher spontaneous polarization along the [111] direction, matching closely to the experimental value. Although large local strains in polycrystalline thin films cannot be ruled out, the $Pca2_1$ phase with moderate interstitial Hf doping serves as a \textit{minimal} model to explain experimental observations. We note that \Hfi~exhibits a negative BEC of $-2e$, indicating that its slight movement induces a change in polarization equivalent to the displacement of two negative elementary charges. This counterintuitive result is likely due to the crowding effect of the confining cage around \Hfi. Nevertheless, since the \Hfi~is strongly confined, it does not change the polarization switching mechanism identified with DFT-based NEB calculations.

\begin{figure}[t]
\includegraphics[width=0.9\textwidth]{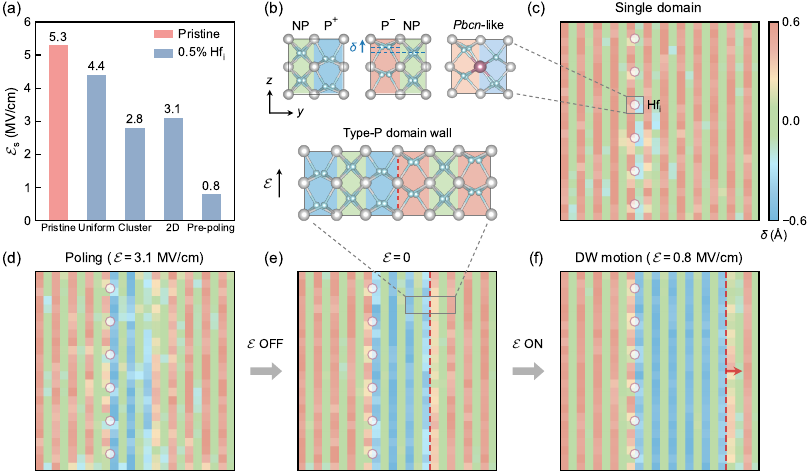}
\caption{Interstitial Hf dopants promoted polarization reversal at elevated temperatures. (a) Switching fields (\Es) of pristine and differently distributed interstitial-doped HfO$_2$. (b) Alternately arranged nonpolar (NP) and polar (P$^-$/P$^+$) oxygen atoms in a $Pca2_1$ unit cell. The P$^-$ oxygen atoms exhibit a positive local displacement ($\delta$) of 0.54~\AA. Distributions of $\delta$ in a 12$\times$12$\times$12 supercell in the presence of two-dimensional-confined interstitial Hf dopants (denoted by hollow circles) between P$^-$ and NP layers without an applied electric field (c), under $\mathcal{E}=3.1$ MV/cm poling (d), with the electric field turned off (e), and under subsequently applied $\mathcal{E}=0.8$ MV/cm (f). The left inset of (c) displays the local $Pbcn$-like structure surrounding an interstitial Hf. The top inset of (e) displays the atomic structure of a mobile type-P domain wall~\cite{Wu24parXiv}.}
\label{fig_md}
\end{figure}

Limited by computational cost, our DFT investigations focused on small supercells. This raises a legitimate question: how does the reduced energy difference between the $O$ and $T$ phases, seen in zero-Kelvin DFT calculations, manifest in larger supercells at elevated temperatures that better represent the experimental sample conditions? To address this, we perform large-scale MD simulations at finite temperatures for an $O$-Hf$_{1.005}$O$_2$ supercell of 20,772 atoms, considering different \Hfi~dopant distributions. The force field is a deep neural network-based model potential~\cite{Ma23p256801}, which accurately reproduces various properties including the energy variation along the switching pathway in Hf$_{1+x}$O$_2$ (see Fig.~S3). All simulations are conducted at 400~K to facilitate the switching process on (sub)nanosecond time scales, thereby mitigating computational cost. We gauge the ease of switching by determining the lowest field strength (\Es) that triggers polarization switching within 200~ps in MD simulations. As shown in Fig.~\ref{fig_md}(a), the switching field is 5.3 MV/cm for pristine single-domain HfO$_2$. Introducing uniformly distributed 0.5\% \Hfi~dopants decreases \Es~to 4.4 MV/cm. The local enrichment of \Hfi~dopants further reduce \Es: the cluster-like and two-dimensional (2D) distributions of \Hfi~lead to \Es~of 2.8 and 3.1 MV/cm, respectively (see dopant distributions in Fig.~S4).

We discover that a simple pre-poling treatment can bring \Es~to an even lower value of $\approx$0.8~MV/cm. The switching process extracted from MD simulations is presented in Fig.~\ref{fig_md}(c-f), with the color scheme explained in Fig.~\ref{fig_md}(b). The starting configuration has 3C oxygen atoms displaced upward (P$^-$) relative to 4C oxygen atom (NP). The 2D-confined \Hfi~dopants are distributed randomly within the $xz$ plane between P$^-$ and NP layers. A main impact of \Hfi~is the slight downward displacement of 4C atoms [left inset of Fig.~\ref{fig_md}(c)]. Upon poling under an electric field of $\mathcal{E}=3.1$ MV/cm, the \Hfi~enriched layer facilitates the nucleation of 180$\degree$ domain walls characterized by oppositely polarized 3C atoms, called type-P walls [see Fig.~\ref{fig_md}(d)]. Locally, the type-P wall resembles a $Pbcn$ unit cell, consistent with DFT calculations showing a spontaneous transformation of the $O$ phase to a $Pbcn$-like structure at high \Hfi~concentration (see Fig.~S2). We confirm that \Hfi~dopants reduce the type-P domain-wall energy (see Fig.~S5). Consequently, polarization switching is driven by the motion of a type-P wall, while the other type-P wall is pinned by \Hfi~dopants. When the external field is turned off, the mobile type-P wall remains stable. After pre-poling, a much lower electric field of 0.8 MV/cm is sufficient to drive the domain-wall motion again.

\begin{figure}[t]
\includegraphics[width=0.9\textwidth]{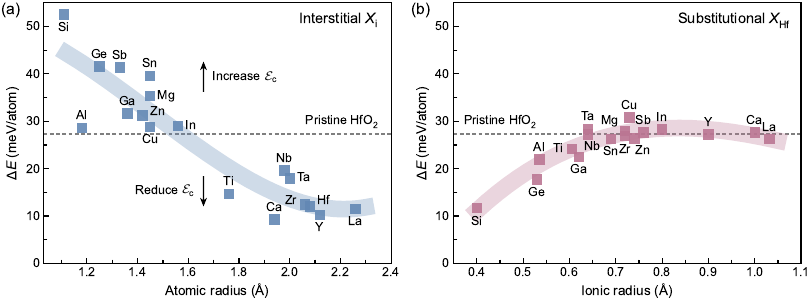}
\caption{Energy difference ($\Delta E= E^T-E^O$) between $T$-phase and $O$-phase HfO$_2$ with different dopants. (a) $\Delta E$ as a function of atomic radius for interstitial $X_{\rm i}$ doping. (b) $\Delta E$ as a function of ionic radius for substitutional $X_{\rm Hf}$ doping. The doping concentration is set to 3\%. The value for pristine HfO$_2$ is denoted by a dashed line as a reference.}
\label{fig_dopant}
\end{figure}

Above results show that the value of $\Delta E$, easily accessible through DFT calculations at the unit cell level, can indeed serve as a useful descriptor for \Ec. To identify more types of dopants for regulating \Ec~in ferroelectric HfO$_2$, we perform high-throughput DFT calculations covering diverse dopant atoms, $X=$ Hf, Zr, Ti, La, Ta, Y, Nb, Sn, Ge, Si, Al, Ga, In, Sb, Mg, Ca, Cu, Zn, with a fixed concentration of 3\%. As shown in Fig.~\ref{fig_dopant}(a), $\Delta E$ between the $T$ and $O$ phases with interstitial dopants ($X_{\rm i}$) is plotted against the dopant atomic radius, revealing a clear negative correlation. Interstitial dopants with large atomic radii, such as Hf, Zr, and Ti, tend to reduce the $\Delta E$, while smaller interstitial atoms such as Si, Ge, and Sn tend to increase $\Delta E$, potentially leading to higher \Ec.

This negative correlation between $\Delta E$ and dopant size for interstitial doping contrasts sharply with substitutional doping, where a volcano-like dependence (mostly a positive correlation) of $\Delta E$ on the ionic radius of the substitutional dopant was observed~\cite{Yang20p064012}. We further examine substitutional doping with atom $X$ ($X_{\rm Hf}$) and plot $\Delta E$ as a function of the ionic radius of $X$ [see Fig.\ref{fig_dopant}(b)]. Different from interstitial doping, the $\Delta E$ of HfO$_2$ under substitutional doping increases with dopant size, consistent with previous DFT results~\cite{Yang20p064012}. Notably, only a few small substitutional dopants, like Si$_{\rm Hf}$, effectively lower $\Delta E$. This has been attributed to the formation of stable $sp^3$ bonds between the substitutional Si dopant and neighboring oxygen atoms in the $T$ phase~\cite{Yang20p064012}.

\begin{figure}[t]
\includegraphics[width=0.5\textwidth]{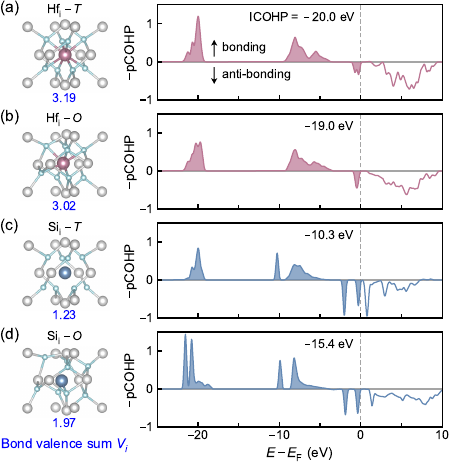}
\caption{Bonding analysis for interstitial Hf and Si with their neighboring oxygen atoms. (a) \Hfi~in $T$ phase. (b) \Hfi~in $O$ phase. (c) \Sii~in $T$ phase. (d) \Sii~in $O$ phase. The left panel shows the local atomic structure surrounding the interstitial atom, with the bond valence sum $V_i=\sum_j\exp{[(R_{\rm o}-R_{ij})/B]}$ highlighted at the bottom. The Brown’s empirical parameters $R_{\rm o}=1.923$~\AA~and $B=0.375$~\AA~for Hf, and $R_{\rm o}=1.624$~\AA~and $B=0.389$~\AA~for Si~\cite{Gagne15p562} are used in the bond valence calculation. The right panel shows the projected crystal orbital Hamilton population (pCOHP) averaged over eight interstitial--oxygen pairs, labeled with the total integration of COHP (ICOHP).}
\label{fig_bond}
\end{figure}

To comprehend the diametrically opposing effects of \Hfi~and \Sii~on $\Delta E$ and \Ec, we analyze the bonding of interstitial atoms in the $T$ and $O$ phases of HfO$_2$. As depicted in Fig.~\ref{fig_bond}, the intercalated atom is surrounded by eight oxygen atoms. We quantify the bonding strength of an interstitial dopant based on the bond-valence conservation principle~\cite{Brown09p6858,Liu13p104102}. According to this principle, each atom $i$ has a preferred atomic valence $V_{0,i}$, which is often the nominal oxidation state of the atom (\eg, $+4$ for Hf). The actual atomic valence, $V_i$, is obtained by summing the individual bond valence ($V_{ij}$) for bonds between the atom and its neighbors ($j$), $V_i=\sum_j V_{ij}$. A smaller deviation of $V_i$ from $V_{0,i}$ reflects stronger overall bonding strength. The bond valence for each oxygen-dopant pair is calculated as $V_{ij}=\exp{[(R_{\rm o}-R_{ij})/B]}$, where $R_{ij}$ is the bond length, $R_{\rm o}$ and $B$ are Brown's empirical parameters~\cite{Gagne15p562}. The calculated $V_i$ for \Hfi~is 3.19 in the $T$ phase and 3.02 in the $O$ phase, indicating stronger bonding of \Hfi~in the $T$ phase. This is further supported by the projected COHP analysis. As shown in the right panel of Fig.~\ref{fig_bond}, the total integration of COHP (ICOHP) for \Hfi$-$O pairs is $-$20.0 and $-$19.0 eV in the $T$ and $O$ phases, respectively. The higher magnitude of ICOHP in the $T$ phase confirms stronger bonding between \Hfi~and oxygen atoms. 

The difference between the energy of the $O$ and $T$ phases containing \Hfi~($E^O[{\rm Hf_i}]$ and $E^T[{\rm Hf_i}]$) can be directly related to the differences in the formation energy of \Hfi~($E_f[{\rm Hf_i}]$) as follows:
\begin{align*}
\Delta E &= E^T[{\rm Hf_i}] - E^O[{\rm Hf_i}] \\
         &= (E^T[{\rm Hf_i}] - E^T_0 -\mu_{\rm Hf}) - (E^O[{\rm Hf_i}] - E^O_0 - \mu_{\rm Hf}) + E^T_0 - E^O_0 \\
         &=E^T_f[{\rm Hf_i}] - E^O_f[{\rm Hf_i}] + \Delta E_0 = \Delta E_f [{\rm Hf_i}] + \Delta E_0,
\end{align*}
where $\mu_{\rm Hf}$ is the chemical potential of Hf and $E_0$ is the energy of the pristine crystal. 
The relatively stronger \Hfi$-$O bonding in the $T$ phase suggests a lower formation energy of \Hfi~than that in the $O$ phase ($E^T_f[{\rm Hf_i}]< E^O_f[{\rm Hf_i}]$)~\cite{Wei21p2104913}, which is responsible for the reduced $\Delta E$ compared to the undoped value ($\Delta E_0$). In the case of \Sii, the pCOHP curve reveals a stronger anti-bonding character, which is corroborated by the small values of $V_i$. Importantly, both the magnitudes of ICOHP and $V_i$ are larger in the $O$ phase than those in the $T$ phase. This points to a lower formation energy of \Sii~in the low-energy $O$ phase, which will further increases the energy difference between $T$ and $O$ phases. 

In summary, our investigations combining both zero-Kelvin DFT calculations and large-scale MD simulations at finite temperatures establish the link between interstitial doping and coercive field in ferroelectric $Pca2_1$ HfO$_2$. Unit-cell-level DFT calculations act as a mean-field-like analysis, demonstrating that the coercive field reduction from interstitial Hf doping is due to the lower defect formation energy in the intermediate $P4_2/nmc$ phase. MD simulations, better representing experimental conditions, suggest that interstitial Hf dopants likely promote switching by forming mobile $Pbcn$-like domain walls. With pre-poling, a switching field of $<$1.0 MV/cm can drive polarization reversal within subnanoseconds. Finally, high-throughput DFT calculations reveal a negative correlation between the switching barrier and the size of the interstitial dopant. The comprehensive understanding of interstitial doping effects in hafnia offers useful guidelines for optimizing the coercive field in this silicon-compatible ferroelectric oxide.

\begin{acknowledgments}
T.Z., L.M. and S.L. acknowledge the supports from National Key R\&D Program of China (2021YFA1202100), National Natural Science Foundation of China (12361141821, 12074319), and Westlake Education Foundation. The computational resource is provided by Westlake HPC Center.
\end{acknowledgments}

\clearpage

\clearpage

\clearpage

\clearpage

\clearpage

\bibliography{SL}

\end{document}